\documentclass[
    a4paper, 
    10pt, 
    twoside, 
]{LTJournalArticle}

\usepackage{indentfirst} 

\usepackage{tabularx}
\usepackage{longtable} 
\usepackage{dblfloatfix} 
\usepackage{float} 
\usepackage{supertabular} 
\usepackage{amsmath} 

\runninghead{Synthetic Student Responses} 

\usepackage{tcolorbox}
\tcbuselibrary{breakable, skins}

\usepackage{xcolor}

\usepackage{enumitem} 

\setlist[enumerate]{noitemsep, topsep=0pt, partopsep=4pt, parsep=4pt}

\definecolor{slate-50}{HTML}{F8FAFC}
\definecolor{slate-100}{HTML}{F1F5F9}
\definecolor{slate-200}{HTML}{E2E8F0}
\definecolor{slate-300}{HTML}{CBD5E1}
\definecolor{slate-400}{HTML}{94A3B8}
\definecolor{slate-500}{HTML}{64748B}
\definecolor{slate-600}{HTML}{475569}
\definecolor{slate-700}{HTML}{334155}
\definecolor{slate-800}{HTML}{1E293B}
\definecolor{slate-900}{HTML}{0F172A}

\newtcolorbox{promptbox}{
    colback=slate-100,
    colframe=slate-400,
    boxrule=0.5pt,
    arc=2pt,
    fontupper=\ttfamily\footnotesize,
    left=10pt,
    right=10pt,
    top=8pt,
    bottom=8pt,
    breakable
}

\newtcolorbox{questionbox}[2][]{%
    enhanced,
    colback=slate-50,
    colframe=slate-400,
    boxrule=0.5pt,
    arc=4pt,
    title={#2},
    fonttitle=\bfseries\color{white},
    attach boxed title to top left={xshift=0.5cm,yshift=-\tcboxedtitleheight/2},
    boxed title style={
        colback=slate-400,
        colframe=slate-400,
        arc=2pt,
        boxrule=0pt,
    },
    top=12pt, 
    breakable,
    #1
}

\newtcolorbox{studentbox}[2][]{%
    enhanced,
    colback=slate-100,
    colframe=slate-500,
    boxrule=0.5pt,
    arc=4pt,
    title={#2},
    fonttitle=\bfseries\color{white},
    attach boxed title to top left={xshift=0.5cm,yshift=-\tcboxedtitleheight/2},
    boxed title style={
        colback=slate-500,
        colframe=slate-500,
        arc=2pt,
        boxrule=0pt,
    },
    top=12pt,
    breakable,
    #1
}

\newtcolorbox{llmbox}[2][]{%
    enhanced,
    colback=slate-200,
    colframe=slate-600,
    boxrule=0.5pt,
    arc=4pt,
    title={#2},
    fonttitle=\bfseries\color{white},
    attach boxed title to top left={xshift=0.5cm,yshift=-\tcboxedtitleheight/2},
    boxed title style={
        colback=slate-600,
        colframe=slate-600,
        arc=2pt,
        boxrule=0pt,
    },
    top=12pt,
    breakable,
    #1
}

\title{Synthetic Student Responses: LLM-Extracted Features for IRT Difficulty Parameter Estimation}

\author{%
    Matías Hoyl\textsuperscript{1} 
}

\date{\footnotesize \textsuperscript{1}School of Education, Stanford University}

\begin{document}

\maketitle

\begin{abstract}
Educational assessment relies heavily on knowing question difficulty, traditionally determined through resource-intensive pre-testing with students. This creates significant barriers for both classroom teachers and assessment developers. We investigate whether Item Response Theory (IRT) difficulty parameters can be accurately estimated without student testing by modeling the response process and explore the relative contribution of different feature types to prediction accuracy. Our approach combines traditional linguistic features with pedagogical insights extracted using Large Language Models (LLMs), including solution step count, cognitive complexity, and potential misconceptions. We implement a two-stage process: first training a neural network to predict how students would respond to questions, then deriving difficulty parameters from these simulated response patterns. Using a dataset of over 250,000 student responses to mathematics questions, our model achieves a Pearson correlation of approximately 0.78 between predicted and actual difficulty parameters on completely unseen questions.
\end{abstract}

\section{Introduction}

Educational assessment plays a vital role in teaching and learning, but creating high-quality questions requires knowing their difficulty level: how challenging they are for students with different abilities. Currently, determining question difficulty requires extensive pre-testing with students followed by statistical analysis, an approach that is effective but creates significant delays and costs in developing new assessments.

For teachers developing classroom assessments, this process is often impractical, leading to tests with unknown psychometric properties. For large-scale assessment programs, the pre-testing requirement consumes substantial resources and limits the speed at which new content can be developed. Both scenarios highlight the need for methods that can predict question difficulty without requiring actual student testing.

This study addresses two specific research questions:
\begin{itemize}
    \item How effectively can we estimate Item Response Theory difficulty parameters without student testing by modeling the student response process?
    \item To what extent do different feature types (LLM-extracted pedagogical features, linguistic characteristics, and semantic embeddings) contribute to the accuracy of difficulty predictions?
\end{itemize}

Rather than treating difficulty prediction as a direct regression problem, which is what most existing research does, we take an approach that mirrors how difficulty emerges in practice: through patterns of student responses.

Our method combines conventional question features with pedagogical insights extracted using Large Language Models (LLMs). These AI tools analyze questions to identify solution steps, required mathematical skills, cognitive complexity, and potential student misconceptions—aspects typically assessed by human experts. The LLM features function much like an experienced teacher would: reading the question and assessing its difficulty based on intuition and experience, combining multiple pedagogical signals to form a holistic judgment. By embedding this pedagogical knowledge in our predictions, we create a system that better reflects how questions function in educational settings.

The remainder of this paper is organized as follows. Section 2 reviews related work on difficulty prediction, highlighting existing approaches and gaps in the literature. Section 3 describes our dataset, which contains over 250,000 student responses to mathematics questions. Section 4 outlines our methodology, including feature engineering and our neural network architecture. Section 5 presents our experimental results, including the primary model's performance, an RMSE-based efficiency evaluation, and a feature ablation study. Finally, Section 6 discusses the implications of our findings, acknowledges limitations, and suggests directions for future research.

\section{Related Work}

\subsection{Traditional Item Difficulty Estimation}

Traditional approaches to difficulty estimation typically rely on either expert judgment or statistical analysis of student responses. Manual calibration by subject matter experts, while common, has been shown to be inconsistent and subjective \cite{rudner2010implementing, impara1998teachers}. Items developed by experts often perform differently than expected when deployed in real settings.

Statistical calibration through pre-testing with actual students provides more reliable estimates but introduces significant delays before questions can be used operationally. This approach typically applies Item Response Theory (IRT) models to estimate difficulty parameters based on student response patterns \cite{rasch1960}. While effective, this process is time-consuming and resource-intensive, creating a bottleneck in the development of high-quality assessments.

The limitations of these traditional methods have motivated research into automated approaches that can predict item difficulty from the question's textual content, potentially reducing or eliminating the need for extensive pre-testing.

\subsection{Text-based Features for Difficulty Prediction}

Research on text-based difficulty prediction has identified several categories of features that contribute to question complexity.

\subsubsection{Linguistic Features}

Linguistic characteristics of questions are strongly associated with their difficulty. Some studies have extracted lexical, syntactic, and semantic features to predict question complexity \cite{alkhuzaey2023text, choi2020predicting}. Common linguistic features include:

\begin{itemize}
    \item \textbf{Lexical features:} Word count, word frequency, word length, and vocabulary difficulty \cite{yaneva2019predicting, benedetto2020framework}
    \item \textbf{Syntactic features:} Sentence length, syntactic complexity, and grammatical structures \cite{choi2020predicting}
    \item \textbf{Semantic features:} Concept density, abstraction level, and semantic similarity between options \cite{bulut2023evaluating}
\end{itemize}

Hickendorff~\cite{hickendorff2013} demonstrated that linguistic factors significantly impact student performance on mathematics assessments, even when controlling for mathematical complexity.

\subsubsection{Domain-specific Features}

Features specific to certain domains have been shown to improve prediction accuracy:

\begin{itemize}
    \item \textbf{Mathematics:} Number of mathematical symbols, presence of graphs or figures, computation complexity \cite{benedetto2020r2de}
    \item \textbf{Reading comprehension:} Text layout, presence of contextual cues, relationship between passages and questions \cite{lawrence2022reading}
    \item \textbf{Multiple-choice questions:} Similarity between options, plausibility of distractors \cite{kurdi2016experimental, yaneva2018automatic}
\end{itemize}

Toyama~\cite{toyama2021makes} emphasized that item difficulty is determined by the interplay of item characteristics, respondent characteristics, and the context in which assessment occurs.

\subsection{Machine Learning Approaches for Difficulty Prediction}

Early approaches to difficulty prediction relied on deterministic methods or simple regression models. Recent advances have introduced increasingly sophisticated machine learning techniques.

\subsubsection{Classical Machine Learning Methods}

Traditional machine learning algorithms have been widely applied to the challenge of difficulty prediction. Linear and logistic regression models have proven effective in establishing relationships between text features and difficulty levels, as demonstrated by Ha et al.~\cite{yaneva2019predicting} in their work on high-stakes medical exams. More sophisticated approaches include the R2DE model proposed by Benedetto et al.~\cite{benedetto2020r2de}, which leverages random forests to simultaneously predict both IRT difficulty and discrimination parameters based on TF-IDF features extracted from question text. The literature also shows that feature-based ensemble methods have achieved notable success across various educational domains, with Ha et al.~\cite{yaneva2019predicting} demonstrating that combining different feature types can enhance prediction accuracy compared to single-feature approaches.

\subsubsection{Neural Network Models}

Deep learning approaches have increasingly been applied to difficulty prediction, with several notable innovations emerging in recent years. Recurrent Neural Networks, particularly those utilizing Long Short-Term Memory (LSTM) architectures, have demonstrated effectiveness in capturing sequential patterns within question text, as shown in the work of Huang et al.~\cite{huang2017question}. Building on this foundation, Qiu et al.~\cite{qiu2019question} introduced a more sophisticated approach with their Document-enhanced Attention Network (DAN), which improves prediction accuracy by enriching questions with relevant domain documents. This contextual enhancement allows the model to better understand the relationship between questions and their subject matter. Further advancing the field, Xue et al.~\cite{xue2020predicting} explored transfer learning techniques, demonstrating that models pre-trained on related tasks such as response time prediction can be effectively fine-tuned for difficulty estimation, leveraging knowledge gained from one educational assessment task to improve performance on another.

\subsubsection{Transformer-based Approaches}

The most recent research has explored transformer-based models for difficulty prediction, leveraging their superior ability to capture contextual relationships. Benedetto et al.~\cite{benedetto2021transformers} compared BERT and DistilBERT models for difficulty estimation, finding that transformer models outperform previous approaches by up to 6.5\% in terms of RMSE. Building on this work, Gombert et al.~\cite{gombert2024predicting} employed scalar-mixed transformer encoders with specialized regression heads, showing significant improvements over baseline models. Further advancing the field, Kapoor et al.~\cite{kapoor2025prediction} incorporated embeddings from various LLMs (ModernBERT, BERT, and LLAMA) alongside linguistic features, achieving a correlation of 0.77 between true and predicted difficulty.

A key advantage of transformer models is their ability to capture complex semantic relationships within text without requiring extensive feature engineering. Aradelli~\cite{aradelli2020transformers} demonstrated that fine-tuning pre-trained transformers on domain-specific corpora further enhances prediction accuracy. These approaches represent the current state-of-the-art in difficulty prediction, combining the contextual understanding capabilities of transformer architectures with specialized training techniques to achieve unprecedented levels of accuracy in estimating question difficulty from text alone.

\subsection{Feature Extraction Using Language Models}

Recent studies have explored the use of Large Language Models (LLMs) not just for prediction but also for feature extraction:

\begin{itemize}
    \item \textbf{Procedural complexity:} Using LLMs to quantify the number of steps required to solve problems \cite{liu2023improving}
    \item \textbf{Skill identification:} Extracting the specific skills required to answer questions \cite{didolkar2024metacognitive}
    \item \textbf{Cognitive level assessment:} Classifying questions according to Bloom's taxonomy \cite{scaria2024automated}
    \item \textbf{Misconception analysis:} Identifying potential student misconceptions associated with questions \cite{sadihin2024proposalml}
\end{itemize}

These approaches leverage the reasoning capabilities of LLMs to extract pedagogically meaningful features that might be difficult to capture through conventional feature engineering approaches.

\subsection{Gaps in Current Research}

Despite significant advances in difficulty prediction methods, two important gaps remain in current research.

First, while transformer-based models have shown remarkable results, often outperforming traditional machine learning approaches by leveraging contextual relationships within question text, there has been limited exploration of using Large Language Models (LLMs) to extract pedagogically meaningful features from questions. Most studies rely on conventional linguistic features or embedding-based approaches, neglecting the potential of LLMs to identify more complex attributes, such as procedural complexity, specific mathematical skills, cognitive levels, and potential misconceptions.

These LLM-extracted features could provide richer signals about question characteristics that are difficult to capture through traditional feature engineering approaches. The few studies that have begun exploring LLM-based feature extraction have shown promising results, but this approach remains underutilized in the difficulty prediction literature.

Second, most current research approaches difficulty prediction as a direct regression problem, where models are trained to predict difficulty parameters directly from question text and features. This direct approach, while intuitive, bypasses a crucial intermediate step: modeling how students with varying ability levels would actually respond to these questions.

Few studies have explored the alternative approach of first simulating student responses to questions and then deriving difficulty parameters from these simulated response patterns. This simulation-based approach offers several advantages over direct prediction. It better mimics the actual process through which question difficulty is determined in real educational settings, allowing difficulty parameters to emerge naturally from student performance patterns rather than being directly predicted. Additionally, it provides a more interpretable model that captures both question characteristics and student abilities, while also producing additional useful metrics beyond difficulty, such as discrimination parameters and expected response patterns.

This gap presents an opportunity to develop more robust difficulty estimation methods. If a model can successfully simulate how students would respond to new questions, it could potentially be applied to any unseen question to estimate difficulty parameters without requiring actual student testing. Such an approach would significantly reduce the resource burden of developing high-quality assessments while maintaining psychometric validity.

The present study addresses both of these gaps by (1) leveraging LLMs to extract complex pedagogical features from questions and (2) developing a neural network model that first predicts student responses and then derives IRT difficulty parameters from these predictions, offering a more comprehensive approach to the difficulty estimation problem.

\section{Data}

The dataset for this study comes from Zapien, an adaptive math learning platform in Chile. It includes 251,851 student answers to 4,696 unique math questions, provided by 1,875 students.

Since the original question text and answer options were in Spanish, we first translated them to English using Google's Gemini 2.0 Flash model that excels at translation. This step was crucial for optimizing the performance of the English-based text embedding models used later in our feature engineering.

Next, we established benchmark difficulty values for each question. Using the entire dataset of 251,851 student responses, we applied a standard 1-Parameter Logistic (1PL) Item Response Theory model. This model estimates a difficulty parameter for each question based on the observed patterns of correct and incorrect answers across all students. These benchmark parameters represent the "ground truth" difficulty derived from real student performance.

\begin{figure}[H]
    \centering
    \includegraphics[width=0.8\columnwidth]{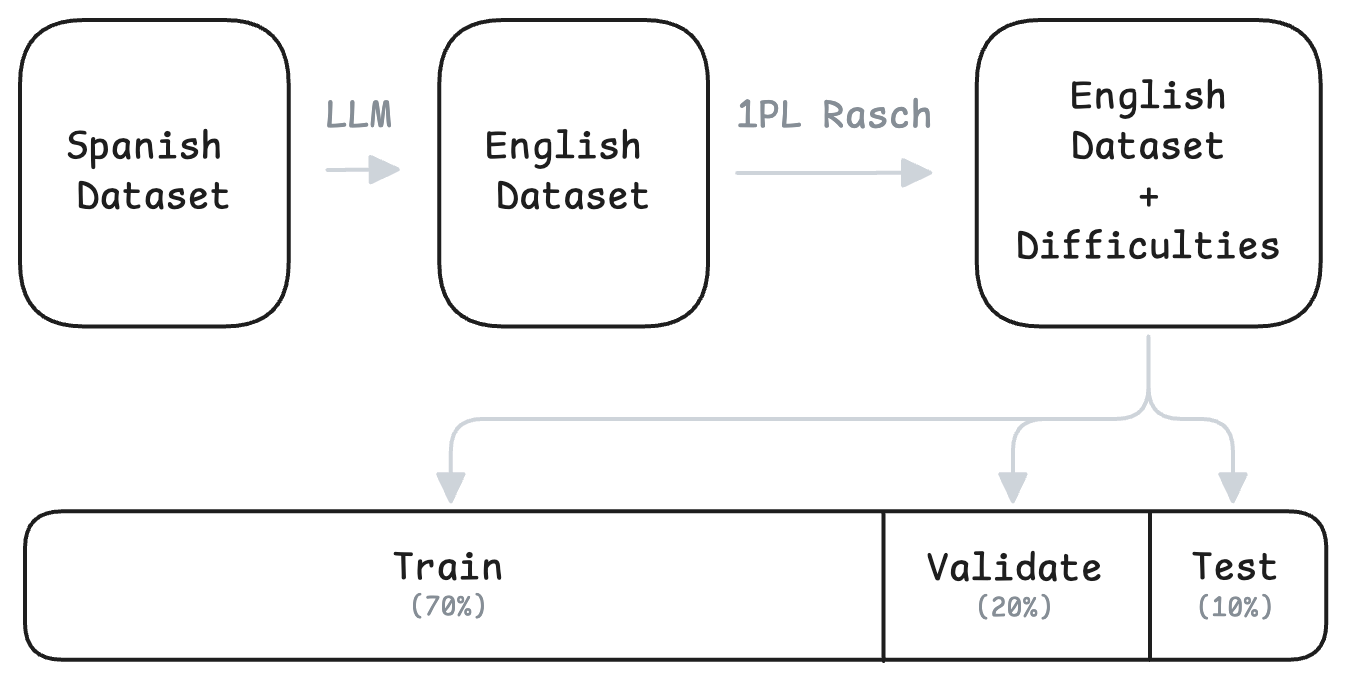}
    \caption{Data processing pipeline.}
    \label{fig:data-pipeline}
\end{figure}

Finally, to train and evaluate our prediction model, we split the 4,696 questions into three sets: training (70\%), validation (20\%), and a holdout test set (10\%). This split was done using stratified sampling based on both the difficulty calculated earlier and the average student correctness for each question. Stratification ensures that all three sets have comparable distributions of question difficulty and student performance patterns.

\begin{table}[H] 
    \centering
    \begin{tabular}{lll}
        \hline
        \textbf{Split} & \textbf{Questions} & \textbf{Answers} \\
        \hline
        Training & 3,286 (70\%) & 178,031 (71\%) \\
        Validation & 940 (20\%) & 49,100 (19\%) \\
        Holdout Test & 470 (10\%) & 24,720 (10\%) \\
        \hline
    \end{tabular}
    \caption{Dataset splits.}
    \label{tab:dataset-splits} 
\end{table}



\section{Methods}

This section describes our methodological approach for predicting question difficulty. We employ a two-stage process where we first train a neural network to predict whether a student will answer a mathematics question correctly. The network processes both student representations and comprehensive question features (including text embeddings, linguistic characteristics, structural properties, answer option features, and LLM-generated pedagogical features detailed in the next section). For each student-question pair, the model outputs a probability of correct response, which, when applied across multiple students and questions, generates a complete prediction matrix of expected student performance.

\begin{figure}[H]
    \centering
    \includegraphics[width=0.8\columnwidth]{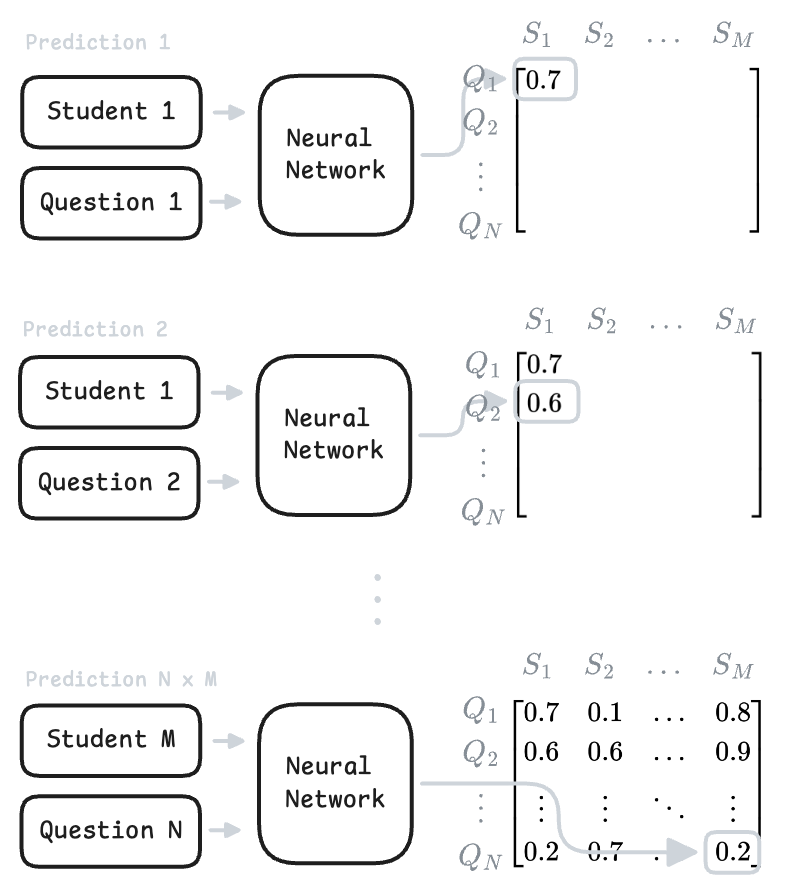}
    \caption{Stage 1: Generation of a correctness prediction matrix using a neural network.}
    \label{fig:stage1}
\end{figure}

Second, we apply a 1-Parameter Logistic (1PL) Item Response Theory model to the complete prediction matrix, using maximum likelihood estimation to derive difficulty parameters for each question. This approach transforms the neural network's correctness predictions into standardized difficulty values that are directly comparable to our benchmark parameters.

\begin{figure}[H]
    \centering
    \includegraphics[width=0.8\columnwidth]{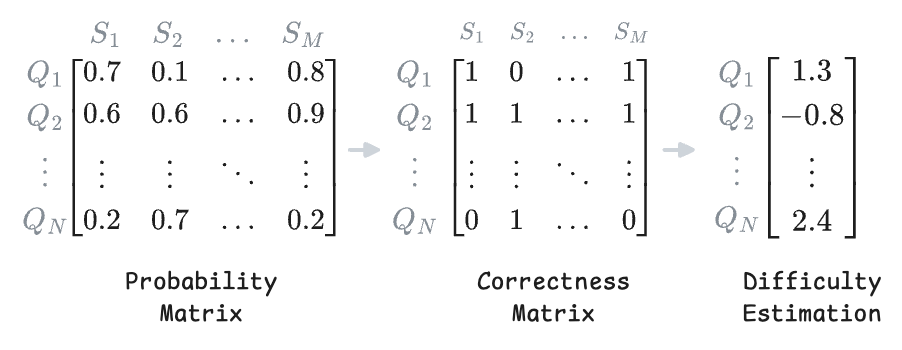}
    \caption{Stage 2: Estimation of IRT difficulty parameters from a correctness prediction matrix.}
    \label{fig:stage2}
\end{figure}

\subsection{Stage 1: Neural Network for Correctness Prediction}

\begin{figure}[H]
    \centering
    \includegraphics[width=0.9\columnwidth]{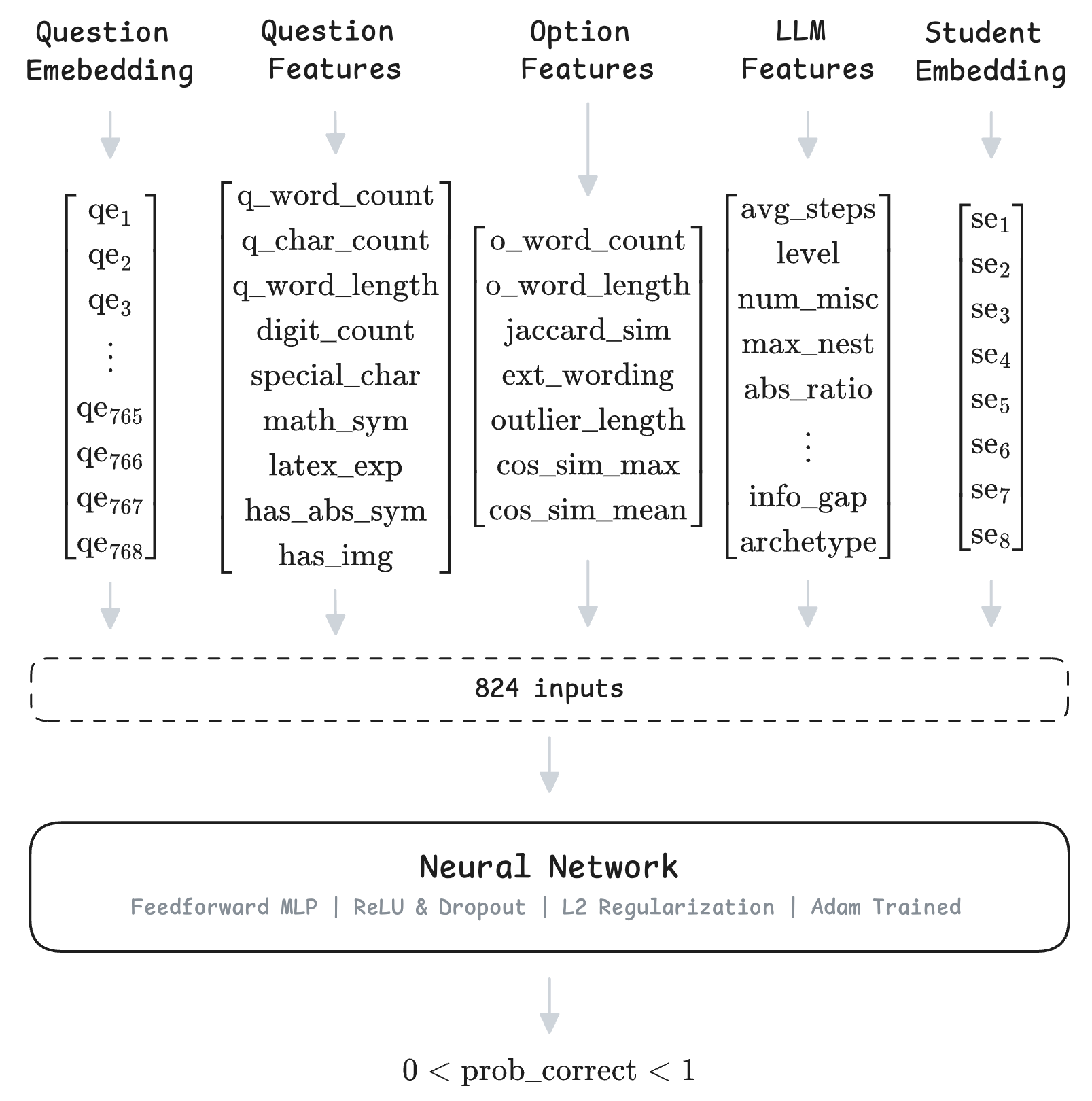}
    \caption{Schematic of the neural network model for student correctness prediction, showing the integration of question embeddings, feature vectors, and student representations.}
    \label{fig:stage1-detailed}
\end{figure}

Our neural network model predicts the probability of a student answering a question correctly by integrating information from five primary sources:

\begin{enumerate}
    \item \textbf{Question Text Embedding}: A semantic vector representing the combined text of the question title and its answer options.
    \item \textbf{Question Features}: Linguistic and structural characteristics derived from the question\'s text (e.g., word count, presence of images or LaTeX).
    \item \textbf{Option Features}: Linguistic and structural characteristics derived from the answer options (e.g., answer length variance, similarity between options).
    \item \textbf{LLM-Generated Features}: Pedagogical insights extracted using Large Language Models (e.g., estimated solution steps, cognitive complexity, potential misconceptions).
    \item \textbf{Student Embedding}: A learned vector representing the student, for capturing student-question interactions.
\end{enumerate}

\subsubsection{Question Text Embeddings}
To capture the semantic meaning of the entire question context, we generated contextual embeddings for the question title combined with its answer options. The text was formatted as:

    \begin{promptbox}
        \textbf{Question:} [problem\_statement]\\
        \textbf{Correct Answer:} [correct\_option]\\
        \textbf{Wrong Answer 1:} [wrong\_option\_1]\\
        ... \\
        \textbf{Wrong Answer 4:} [wrong\_option\_4]
    \end{promptbox}
    
We fed this formatted text into the ModernBERT transformer model (\texttt{nomic-ai/modernbert-embed-base}) \cite{warner2024smarterbetterfasterlonger}. ModernBERT was selected due to its advantages over traditional transformer architectures, including a longer context window, superior performance on benchmark tasks such as GLUE, and increased efficiency compared to models like RoBERTa and DeBERTa. This model generates contextual representations from its internal layers for the input text. We then averaged these representations across the entire input (mean pooling) to create a single 768-dimensional vector (embedding) capturing the overall meaning. Finally, this embedding was L2 normalized—scaling it to unit length to ensure consistent input magnitudes for the network—and used directly as an input to our neural network.

\subsubsection{Question Features}
These features capture characteristics of the question\'s text and structure, independent of the answer options:
\begin{itemize}
    \item \textbf{Basic Text Metrics:} Word count, character count, average word length.
    \item \textbf{Content Type Metrics:} Count of digits, special characters, mathematical symbols, and LaTeX expressions.
    \item \textbf{Abstractness \& Structure:} Indicator for abstract symbols and presence of an image.
\end{itemize}
These features provide a basic profile of the question's textual complexity and format.

\subsubsection{Option Features}
These features focus on the characteristics of the answer options and their relationships:
\begin{itemize}
    \item \textbf{Option Text Metrics:} Average option length, average option word count, variance in option lengths.
    \item \textbf{Inter-Option Similarity:} Standard deviation of Jaccard similarity between option texts.
    \item \textbf{Option Structure:} Indicators for options with extreme wording like "None/All of the above" or outlier lengths.
    \item \textbf{Embedding-Derived Metrics:} The average cosine similarity between the correct option and distractors, and the average distance between distractor embeddings. These capture semantic relationships between the answer choices.
\end{itemize}
These features describe the choice set presented to the student, including how similar or distinct the options are, both textually and semantically.

\subsubsection{Feature Extraction Using LLMs}
We used Google's Gemini models (primarily \texttt{gemini-2.0-flash} for simpler feature extractions and \texttt{gemini-2.5-flash} or \texttt{gemini-2.5-pro} for more complex ones like archetype classification, as detailed in Appendix \ref{app:llm_prompts}) to extract pedagogically relevant features from the question text and, where applicable, provided solutions. To ensure stability and mitigate the probabilistic nature of LLMs, features were typically generated over N runs (e.g., 3 runs for numerical estimates like information gap, or 1 run for simpler classifications) and then aggregated (e.g., by averaging numerical values or using the mode for categorical ones). The prompts used aimed to elicit chain-of-thought reasoning from the LLM before providing the final structured JSON output.

\paragraph{Procedural Complexity and Cognitive Demand}
\begin{itemize}
    \item \textbf{Number of Steps}: The average number of discrete, pedagogically atomic solution steps required to solve the question. The prompt guided the LLM to break down solutions in a manner appropriate for student instruction (see Appendix \ref{app:prompt_avg_steps} for a conceptual prompt example).
    \item \textbf{Bloom's Taxonomy Level}: The primary cognitive level assessed by the question, based on Bloom's Taxonomy (1-6, see Appendix \ref{app:bloom_taxonomy}).
    \item \textbf{Number of Misconceptions}: The average number of potential student misconceptions associated with the question. The prompt asked for an exhaustive, atomic list (see an example in Appendix \ref{app:num_misconceptions}).
\end{itemize}

\paragraph{Mathematical and Symbolic Content Features}
\begin{itemize}
    \item \textbf{Max Expression Nesting Depth}: The maximum nesting depth of mathematical expressions (e.g., parentheses, functions, fractions) found in the question (see Appendix \ref{app:prompt_nesting_depth}).
    \item \textbf{Ratio Abstract Concrete Symbols}: The ratio of abstract symbols (variables) to concrete numerical values in the question (see Appendix \ref{app:prompt_abstract_ratio}).
    \item \textbf{Units Check}: Binary indicator (1/0) if checking or converting units is significantly involved (see Appendix \ref{app:prompt_units_check}).
\end{itemize}

\paragraph{Question-Answer Relationship and Context Features}
\begin{itemize}
    \item \textbf{Question Answer Info Gap}: An LLM-estimated measure (1-4 scale) of the information gap between the question stem and the knowledge needed to derive the answer (see Appendix \ref{app:prompt_info_gap}).
    \item \textbf{Real World Context Flag}: Binary indicator (1/0) if the question is set in a real-world context versus purely abstract, determined by LLM analysis (see Appendix \ref{app:prompt_realworld_flag}).
\end{itemize}

\paragraph{Distractor Plausibility Features}
\begin{itemize}
    \item \textbf{Distractor Plausibility (Max)}: The average (over 3 runs) of the maximum plausibility scores (1-5 scale) assigned by an LLM among the distractor options for a given question (see Appendix \ref{app:prompt_distractor_plausibility}).
    \item \textbf{Distractor Plausibility (Mean)}: The average (over 3 runs) of the mean plausibility scores (1-5 scale) assigned by an LLM to the distractor options for a given question (see Appendix \ref{app:prompt_distractor_plausibility}).
    \item \textbf{Extreme Wording Option Count}: Count of options containing extreme wording (e.g., "always", "never").
\end{itemize}

\paragraph{Categorical Pedagogical Features}

The following LLM-derived features were treated as categorical and subsequently one-hot encoded before being fed into the neural network:

\begin{itemize}
    \item \textbf{Knowledge Dimension}: The primary dimension of knowledge assessed (Factual, Conceptual, or Procedural) (see Appendix \ref{app:prompt_knowledge_dimension}).
    \item \textbf{Most Complex Number Type}: The highest level of numerical complexity (1-5 scale: Integer to Complex Number) present in the question or options (see Appendix \ref{app:prompt_complex_number_type}).
    \item \textbf{Problem Archetype}: The general type or structure of the problem (e.g., Word Problem - Calculation, Equation Solving, Data Interpretation, etc.) (see Appendix \ref{app:prompt_problem_archetype}).
\end{itemize}

\subsubsection{Student Embeddings}

To account for individual student abilities, our neural network incorporates a dedicated user embedding layer. This layer maps each unique student ID to a dense, learnable 8-dimensional vector. These embeddings allow the model to capture latent aspects of each student's proficiency from the data, effectively learning how a particular student might interact with and respond to a specific question. Incorporating these student-specific representations is crucial. Once the network has learned these student ability profiles from the training data (which involved approximately 1,800 unique students), it becomes highly flexible. We can then introduce any new mathematics question (represented by its own features) to the trained model. The model, leveraging the learned student embeddings, can then predict how each of those ~1,800 students would likely answer that new question. This process generates a full matrix of "synthetic student responses" for the new item without requiring actual student testing. These simulated responses, in turn, support a more robust estimation of the question's difficulty.

\subsubsection{Neural Network Arquitecture}

The network's layer-by-layer structure is as follows:

\begin{itemize}
    \item The user ID is passed through an embedding layer (8-dimensional output), followed by flattening and a dropout layer (rate 0.25).
    \item The 48 numerical features are processed by a dense layer (32 units, ReLU activation, L2 regularization of 0.0005), followed by a dropout layer (rate 0.25).
    \item The 768-dimensional question text embedding is processed by a dense layer (32 units, ReLU activation, L2 regularization of 0.0005), followed by a dropout layer (rate 0.25).
    \item The outputs of these three pathways are concatenated.
    \item This concatenated vector is then passed through two shared dense layers (64 units and then 32 units, both with ReLU activation, L2 regularization of 0.0005, and dropout rate of 0.25 after each).
    \item The final layer is a single dense unit with a sigmoid activation function to output the probability of a correct answer.
\end{itemize}

The specific hyperparameters for training and other configuration details of this model are summarized in Appendix \ref{app:nn_config_details}.

\subsection{Stage 2: IRT Parameter Estimation from Predictions}

After training the neural network on 70\% of the questions and validating on 20\%, we used the model to generate predictions for the 10\% completely unseen questions in the holdout test set. For each of the 470 holdout questions, we generated predictions for all unique students present in the training/validation data. We first preserved the raw probability outputs from the model, representing each student's likelihood of answering correctly.

To prepare these outputs for IRT modeling, we applied a 0.5 threshold to the probabilities to convert them to binary predictions (0 for incorrect, 1 for correct). This binarization step is crucial as it simulates the actual outcome of a student attempting a question—they either answer it correctly or incorrectly, rather than achieving a probabilistic score. This conversion thus produced a dichotomous correctness matrix where each cell contained either 0 or 1, mirroring real-world student response data.

We then used these binarized correctness matrices as inputs to an Item Response Theory estimation process. We implemented a 1-Parameter Logistic (1PL) IRT model to estimate question difficulty parameters. The 1PL model is defined as:
\begin{equation}
P(X_{ij}=1 | \theta_j, \beta_i) = \frac{e^{(\theta_j - \beta_i)}}{1 + e^{(\theta_j - \beta_i)}}
\end{equation}
where $P(X_{ij}=1)$ is the probability of student $j$ answering question $i$ correctly, $\theta_j$ represents the ability of student $j$, and $\beta_i$ represents the difficulty of question $i$. In our application, the $\theta_j$ parameters are estimated based on the patterns in the binarized \textit{predicted} responses.
The parameter estimation employed a maximum likelihood approach using TensorFlow.

\section{Results}

\subsection{Model Performance}

Our neural network model achieved good performance in predicting student correctness. The following metrics were obtained on the holdout test set (470 questions):

\begin{table}[H]
    \centering
    \begin{tabular}{lc}
        \hline
        \textbf{Metric} & \textbf{Value} \\
        \hline
        Accuracy & 0.7529 \\
        AUC-ROC & 0.7789 \\
        Precision & 0.8010 \\
        Recall & 0.8719 \\
        F1 Score & 0.8350 \\
        \hline
    \end{tabular}
    \caption{Classification performance metrics on the holdout test set.}
    \label{tab:classification-metrics-test}
\end{table}

These metrics indicate the model's robust ability to predict student responses on unseen questions, providing a good foundation for the subsequent IRT parameter estimation from these predictions.

\subsection{IRT Parameter Estimation}

Our primary objective was to estimate IRT difficulty parameters from the neural network's binarized predictions on the holdout test set. The alignment between these NN-derived difficulty parameters and the benchmark difficulty parameters (calculated from the full original student response dataset) is shown below:

\begin{table}[H]
    \centering
    \begin{tabular}{lc}
        \hline
        \textbf{Metric} & \textbf{Value} \\
        \hline
        Pearson correlation & 0.7791 \\
        Spearman rank correlation & 0.7606 \\
        RMSE & 1.1857 \\
        Number of common questions & 470 \\
        \hline
    \end{tabular}
    \caption{IRT difficulty parameter estimation metrics on the holdout test set.}
    \label{tab:irt-metrics-1pl}
\end{table}

The high correlation values demonstrate a strong positive relationship between the difficulty parameters estimated via our modeling approach and those derived from actual student response data.

\begin{figure}[H]
    \centering
    \includegraphics[width=0.8\columnwidth]{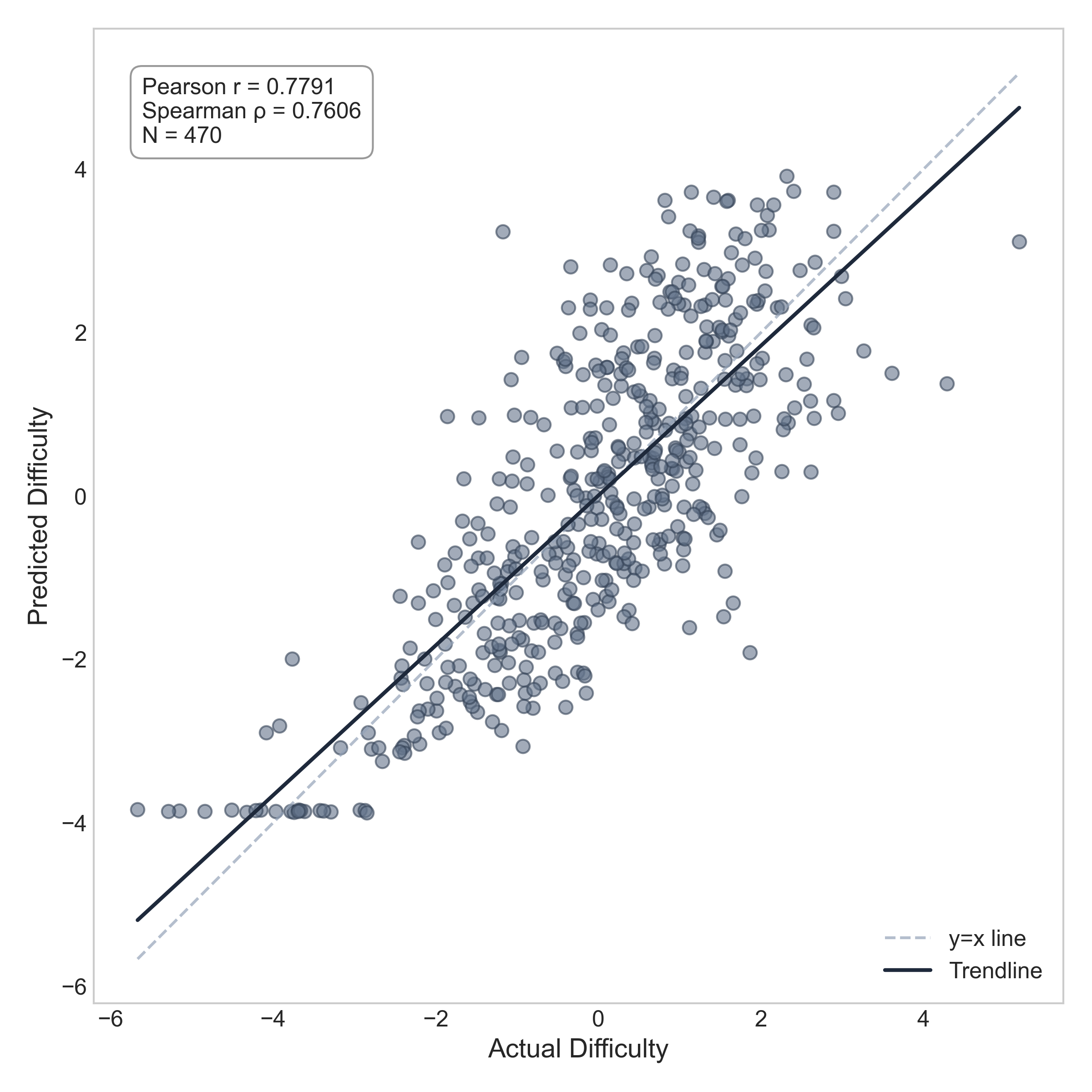} 
    \caption{Correlation between benchmark 1PL IRT difficulty and NN-derived  difficulty on the holdout test set.}
    \label{fig:difficulty-correlation-1pl}
\end{figure}

\begin{figure}[H]
    \centering
    \includegraphics[width=0.8\columnwidth]{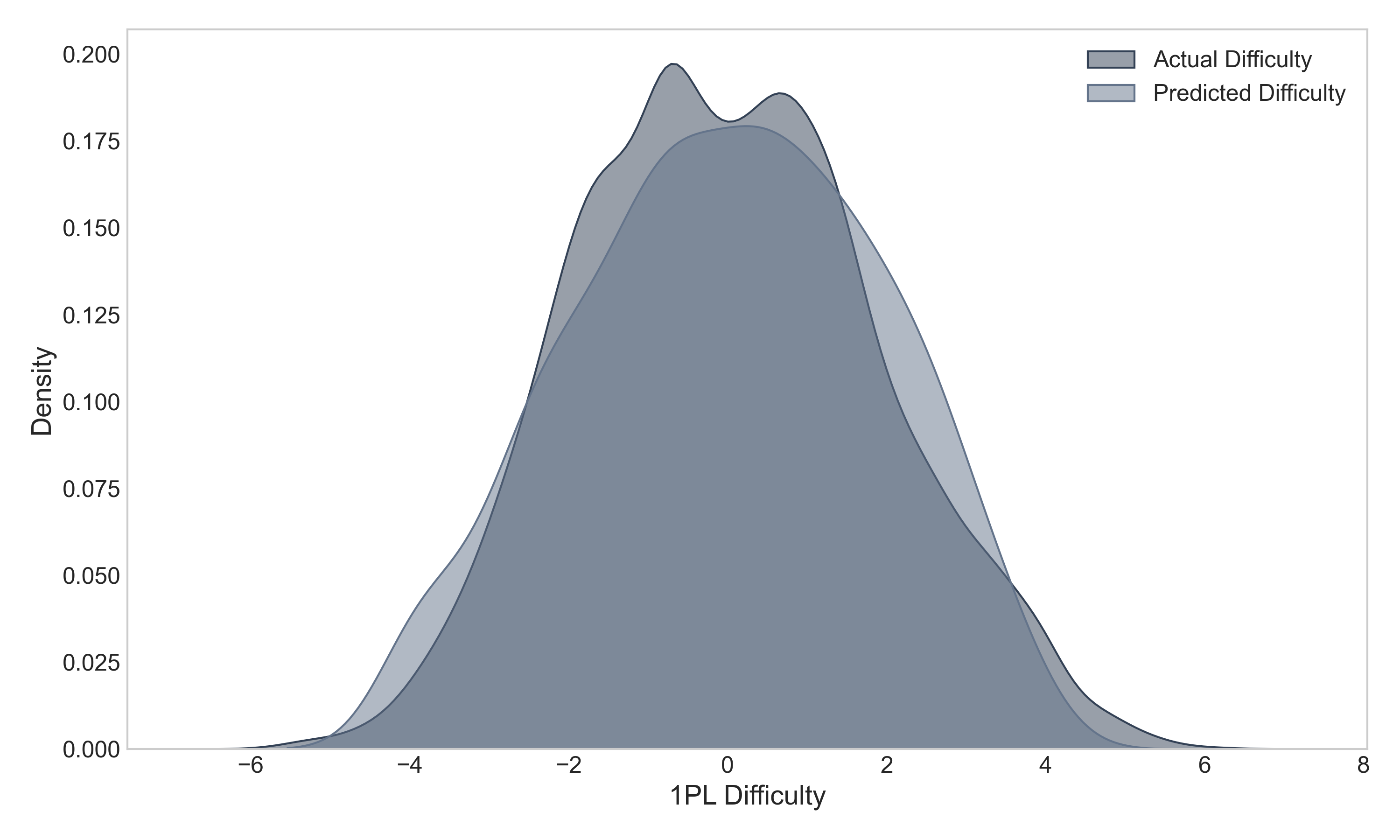} 
    \caption{Distribution comparison of benchmark difficulty and NN-derived difficulty for the holdout test set.}
    \label{fig:difficulty-distribution-1pl}
\end{figure}

As shown in Figures \ref{fig:difficulty-correlation-1pl} and \ref{fig:difficulty-distribution-1pl}, our model captures both the relative ordering and, to a good extent, the distribution of question difficulties.

\subsection{RMSE-Based Efficiency Evaluation}

To understand the practical value of our model, we compared its performance in estimating question difficulty to traditional methods that rely on actual student data. Specifically, we wanted to quantify how many real student responses would be needed in a standard Item Response Theory (IRT) analysis to achieve the same level of accuracy as our model, which uses no new student data for the holdout questions.

For this evaluation, we first established a "ground truth" set of difficulty parameters. These were calculated using a 1-Parameter Logistic (1PL) IRT model applied to \textit{all} available real student answers for the 470 questions in our holdout test set. This represents the most accurate difficulty estimation possible with our dataset for these items.

Next, we took the difficulty parameters predicted by our neural network model for these same 470 holdout questions. The Root Mean Square Error (RMSE) between our model's predicted difficulties and the ground truth was \textbf{1.1857}.

The core of the evaluation involved simulating traditional 1PL IRT analyses using different-sized subsamples of real student responses for the holdout questions. We progressively increased the percentage of real student data used (from 1\% up to 100\%), calculated the difficulty parameters for each subsample, and then computed the RMSE against the ground truth difficulties. To ensure stable estimates, this process of subsampling and RMSE calculation was repeated 10 times for each data percentage, and the average RMSE was used.

\begin{figure}[H]
    \centering
    \includegraphics[width=1\columnwidth]{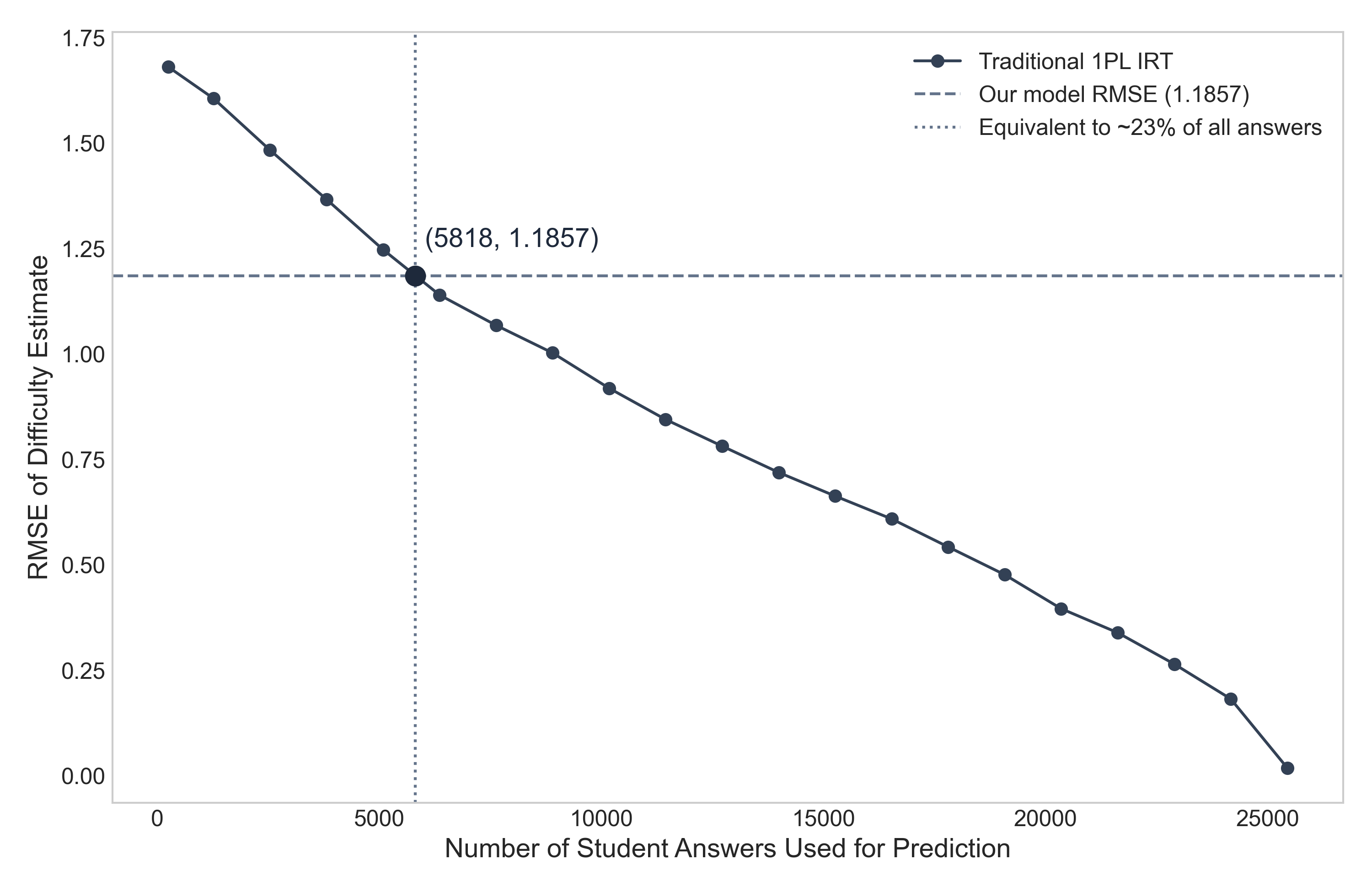}
    \caption{RMSE-Based Efficiency of NN-Derived Difficulty Estimates compared to Traditional 1PL IRT using varying amounts of real student data. As expected, the RMSE decreases as more real student data is used.}
    \label{fig:rmse-efficiency}
\end{figure}

Figure \ref{fig:rmse-efficiency} plots the RMSE achieved by these traditional 1PL IRT estimations as the amount of real student data increases. Our model's RMSE (1.1857) is marked on this plot for comparison. 

This comparison reveals that our model's difficulty estimates achieve an accuracy comparable to what a traditional 1PL IRT analysis would achieve if it used approximately \textbf{5,818 real student answers} for these 470 holdout questions. This number of answers corresponds to roughly \textbf{22.87\%} of the total real student responses available for these particular items in our dataset.

This finding suggests that our approach can estimate question difficulty with an accuracy that would typically require a substantial number of real student interactions. This highlights the potential of our model to make the assessment development process more efficient, reducing the reliance on extensive pre-testing with students while still achieving robust difficulty estimates.

\subsection{Feature Ablation Study}
To address our second research question (understanding how different types of features contribute to predicting question difficulty), we conducted a feature ablation study. In this study, we systematically built models by adding layers of features. We started with a model using only text embeddings and then progressively incorporated traditional linguistic and structural features (derived from both question statements and answer options), and finally added our LLM-extracted pedagogical features. 

Each of these models used the same neural network architecture and IRT comparison framework as our primary reported model. This layered approach allowed us to isolate and observe the impact of each feature group on the accuracy of the estimated IRT difficulty parameters.

The four models we built are described below:

\begin{enumerate}
    \item \textbf{Model 1 (Embeddings Only)}
    Utilized only the \texttt{ModernBert} text embeddings of the formatted question text and options.
    \item \textbf{Model 2 (Embeddings + Question Features)}
    Added linguistic and structural features derived from the question problem statement (e.g., \texttt{word\_count}, \texttt{has\_image}, \texttt{has\_abstract\_symbols}).
    \item \textbf{Model 3 (Embeddings + Question + Option Features)}
    Added linguistic and structural features derived from the answer options (e.g., \texttt{jaccard\_similarity\_std}, \texttt{avg\_option\_length}, \texttt{has\_noneall\_option}, \texttt{cos\_sim\_mean}).
    \item \textbf{Model 4 (Embeddings + Question + Option + LLM Features)}
    Added LLM-derived pedagogical features, both numerical (e.g., \texttt{avg\_steps}, \texttt{level}, \texttt{num\_misconceptions}) and categorical (e.g., \texttt{knowledge\_dimension}, \texttt{problem\_archetype}).
\end{enumerate}

Table \ref{tab:ablation-results} summarizes the performance of these four models on the test set.

\begin{table}[H]
    \centering
    \resizebox{\columnwidth}{!}{
    \begin{tabular}{lcc}
        \hline
        \textbf{Model Configuration} & \textbf{Pearson} & \textbf{Spearman} \\
        \hline
        Embeddings Only & 0.6822 & 0.7064 \\
        Emb. + Question Feat. & 0.6842 & 0.7065 \\
        Emb. + Q \& Opt. Feat. & 0.6758 & 0.7044 \\
        Full Model (All + LLM Feat.) & 0.7791 & 0.7606 \\
        \hline
    \end{tabular}
    }
    \caption{Feature ablation study results: difficulty correlation metrics.}
    \label{tab:ablation-results}
\end{table}

The results of this feature ablation study, presented in Table \ref{tab:ablation-results}, highlight the distinct contributions of each feature group to the estimation of IRT difficulty parameters:

\begin{itemize}
    \item \textbf{Embeddings as a Foundation:} Using only text embeddings (Model 1) provided a strong starting point. This model achieved a Pearson correlation of approximately 0.68 and a Spearman correlation of approximately 0.71 between its predicted difficulty parameters and the benchmark parameters. This indicates that the semantic meaning captured by embeddings alone can account for a substantial part of what makes a question difficult.

    \item \textbf{Minimal Impact from Basic Question and Option Features:} Adding traditional linguistic and structural features derived from the question text (Model 2) offered virtually no improvement in difficulty correlation. Further incorporating similar features from the answer options (Model 3) actually resulted in a slight decrease in Pearson correlation (to around 0.68), while the Spearman correlation remained largely unchanged. This suggests that, once the rich semantic information from embeddings is available, these more basic textual and structural characteristics do not add much more predictive power for question difficulty.

    \item \textbf{Significant Improvement from LLM-Extracted Features:} The most substantial gain in performance came with the addition of LLM-derived pedagogical features (Model 4, the Full Model). This model, which includes all feature types, reached a Pearson correlation of approximately 0.78 and a Spearman correlation of approximately 0.76. This is a marked improvement over the models that relied only on embeddings and traditional features.
\end{itemize}

These findings underscore the value of the LLM-extracted pedagogical features. While text embeddings effectively capture the general meaning of a question and its options, and traditional features describe surface-level text properties, the LLM-derived features seem to tap into deeper aspects related to cognitive load, the complexity of the solution process, and common student challenges. The significant jump in performance when these LLM features were included suggests that these AI-generated pedagogical insights are relevant to more accurately estimating question difficulty without relying on actual student test data.

\section{Discussion}

\subsection{Conclusions}

Our research demonstrates that Item Response Theory difficulty parameters can be accurately estimated without traditional student testing by combining linguistic features with pedagogical insights extracted using Large Language Models. The final model achieved a Pearson correlation of approximately 0.78 (Spearman correlation \textasciitilde0.76) between its predicted 1PL IRT difficulty parameters and benchmark 1PL parameters derived from actual student data, when evaluated on completely unseen questions. This approach addresses a significant challenge in educational assessment: the need for efficient methods to determine question difficulty that don't rely on resource-intensive pre-testing.

The strong performance of our neural network model in predicting student correctness (test set F1 score of 0.8350, AUC of 0.7789) confirms that our feature engineering approach effectively captures many factors influencing student performance. By integrating traditional linguistic and structural features, semantic embeddings of question text, and LLM-extracted pedagogical insights (such as solution step count, cognitive level, and misconception count), we created a rich representation of question characteristics.

Most importantly, the RMSE-based efficiency evaluation showed that our NN-derived 1PL difficulty estimates (RMSE of 1.1857) are comparable in accuracy to using approximately 5,800 real student answers (representing about 22\% of the available data for the holdout questions) in a traditional 1PL IRT estimation. This highlights a substantial potential for data efficiency and cost reduction in assessment development.

These findings suggest that our simulation-based approach can reliably estimate question difficulty, potentially accelerating assessment development while maintaining psychometric quality, and opens new avenues for AI-assisted educational assessment design.

\subsection{Limitations}

Despite the promising results, several limitations of our study should be acknowledged:

\paragraph{Generalizability}: Our dataset originates from Chile, which may limit the generalizability of our findings to student populations elsewhere. Educational systems, curricula, and pedagogical approaches vary across countries and cultures, potentially affecting how question features relate to difficulty in different contexts.
    
\paragraph{LLM Stability}: Large Language Models are inherently probabilistic. Even though we attempted to ensure stability by making multiple LLM calls and aggregating results for features like `avg\_steps` and `num\_misconceptions`, each new call could potentially yield different outputs. This variability introduces some uncertainty into the LLM-extracted features, which could affect the consistency of our predictions.
    
\paragraph{Model Interpretability}: The neural network we employed is essentially a black box that performs well on the prediction task but offers limited insight into the relative importance of different features beyond what was shown in the ablation study. This makes it challenging to draw definitive conclusions about which specific question characteristics contribute most significantly to difficulty within the combined model.
    
\paragraph{Model Sensitivity}: Additionally, our results may vary depending on which LLM vendor we use (e.g., OpenAI, Anthropic, Google) and which inference parameters we select (e.g., temperature, top-p sampling). These choices can significantly impact the quality and consistency of the LLM-extracted features that feed into our model.

\section{Acknowledgments}

I would like to thank my advisor Ben Domingue for his valuable feedback during the writing of this paper and for contributing important ideas like the RMSE difficulty estimation approach and other insights that made the paper clearer and more rigorous.

I am grateful to my program director Sanne Smith for her support and for accompanying me throughout the idea brainstorming process, helping me develop and refine this research concept and pursue it through to completion.

I also thank my fellow MS Education Data Science classmates for all the valuable feedback they provided on early versions of this manuscript and for their support throughout the entire research process.

\bibliographystyle{unsrt}
\bibliography{references}

\clearpage 

\appendix 

\renewcommand{\thesection}{} 
\renewcommand{\thesubsection}{\Alph{subsection}} 

\titleformat{\section}
  [block]
  {\Large\bfseries\centering}
  {}
  {0em}
  {}

\section{Appendix}

\subsection{Step by Step Example}
\label{app:prompt_avg_steps}
The following is an example prompt used to generate step by step solution explanations for a mathematical problem:

\begin{questionbox}{Prompt}
    You are an expert in math pedagogy. Your task is to answer the following question deconstructing it to the most elemental steps. Don't skip any step. Don't assume anything about the reader, try to solve it in the most atomic and pedagogical way possible.

    \vspace{1em}
  
    Here is the question:
    \vspace{1em}

    Question: What is the decreasing order of the roots $4\sqrt{3}$, $3\sqrt{4}$, $2\sqrt{5}$, and $5\sqrt{2}$?  

    A) $2\sqrt5 > 4\sqrt3 > 5\sqrt2 > 3\sqrt4$

    B) $2\sqrt5 > 3\sqrt4 > 4\sqrt3 > 5\sqrt2$

    C) $5\sqrt2 > 3\sqrt4 > 4\sqrt3 > 2\sqrt5$

    D) $5\sqrt2 > 4\sqrt3 > 3\sqrt4 > 2\sqrt5$

    E) $2\sqrt5 > 4\sqrt3 > 5\sqrt2 > 3\sqrt4$

    Correct Answer: D

    \vspace{1em}

    First, think about your task step by step inside <thinking></thinking> tags. Your thinking process should try to go over all the posible ways to solve the question and try to find the most atomic way to solve it.

    \vspace{1em}

    Then, when you are ready to answer, after your </thinking> tag you should write your response using "Step" as the key and the step as the value. For example:

    \vspace{1em}

    <thinking>

      Thinking process... Think as long as you need to.

    </thinking>

    \vspace{1em}

    Step 1: ...

    Step 2: ...

    Step 3: ...

    ...

    \vspace{1em}

    Important instructions:

    - Don't skip any step.
    
    - Don't assume anything about the reader, try to solve it in the most atomic and pedagogical way possible.

    - Use as many steps as you need to.
\end{questionbox}

And here is the LLM's response:

\begin{studentbox}{Thinking Process}
    <thinking>

    Okay, let's break down this problem. We need to compare four numbers:  $4\sqrt{3}$, $3\sqrt{4}$, $2\sqrt{5}$, and $5\sqrt{2}$ and arrange them in decreasing order. The main challenge is that the numbers are in a mixed form (integer multiplied by a square root). To compare them easily, we need to get rid of the square roots or the integers outside the square root.

    \vspace{1em}

    The easiest way to compare these values is to bring the coefficients *inside* the square root.

    \vspace{1em}

    Let's consider how to bring a number inside a square root. For example, consider $a\sqrt{b}$. We can rewrite this as $\sqrt{a^2 * b}$.

    \vspace{1em}

    So, the first step is to convert all the numbers into the form of a single square root.

    \vspace{1em}

    Then, we simply compare the numbers *inside* the square roots, because if $a > b$, then $\sqrt{a} > \sqrt{b}$.

    \vspace{1em}

    Finally, we arrange the original numbers based on their corresponding squared values, in decreasing order.

    \vspace{1em}

    Let's start the solution.

    </thinking>

    \vspace{1em}
\end{studentbox}

\begin{llmbox}{Response}
    Step 1: Convert $4\sqrt{3}$ into the form $\sqrt{x}$.

    We have $4\sqrt{3} = \sqrt{4^2 * 3} = \sqrt{16 * 3} = \sqrt{48}$
    
    \vspace{1em}

    Step 2: Convert $3\sqrt{4}$ into the form $\sqrt{x}$.

    We have $3\sqrt{4} = \sqrt{3^2 * 4} = \sqrt{9 * 4} = \sqrt{36}$
    
    \vspace{1em}

    Step 3: Convert $2\sqrt{5}$ into the form $\sqrt{x}$.

    We have $2\sqrt{5} = \sqrt{2^2 * 5} = \sqrt{4 * 5} = \sqrt{20}$
    
    \vspace{1em}

    Step 4: Convert $5\sqrt{2}$ into the form $\sqrt{x}$.

    We have $5\sqrt{2} = \sqrt{5^2 * 2} = \sqrt{25 * 2} = \sqrt{50}$
    
    \vspace{1em}

    Step 5: Now we have $\sqrt{48}$, $\sqrt{36}$, $\sqrt{20}$, and $\sqrt{50}$. We can easily compare the values inside the square roots: 48, 36, 20, and 50.

    \vspace{1em}

    Step 6: Arrange the values inside the square roots in decreasing order:

    50 > 48 > 36 > 20

    \vspace{1em}

    Step 7: Replace the values inside the square roots with their original forms:

    $5\sqrt{2} > 4\sqrt{3} > 3\sqrt{4} > 2\sqrt{5}$

    \vspace{1em}

    Step 8: Compare this result with the given options.

    The correct option is D: $5\sqrt{2} > 4\sqrt{3} > 3\sqrt{4} > 2\sqrt{5}$.
\end{llmbox}

\subsection{Bloom's Cognitive Level Table}
\label{app:bloom_taxonomy}
Table \ref{tab:blooms-taxonomy} outlines Bloom's Taxonomy levels used to classify the cognitive demands of problems. Each level is an ordinal number in our feature encoding. The LLM was provided with this rubric and asked to determine the primary cognitive level for each question (feature: \texttt{level}).
\begin{table}[H]
    \centering
    \begin{tabular}{p{0.30\columnwidth}p{0.60\columnwidth}}
        \hline
        \textbf{Cognitive Level} & \textbf{Description} \\
        \hline
        Remember & Recall facts and basic concepts \\
        Understand & Explain ideas or concepts, interpret, summarize \\
        Apply & Use information in new situations, execute procedures \\
        Analyze & Draw connections among ideas, break down into parts \\
        Evaluate & Justify a stand or decision, verify, critique \\
        Create & Produce new or original work, design, construct \\
        \hline
    \end{tabular}
    \caption{Bloom's Taxonomy of cognitive levels applied to mathematical problem classification.}
    \label{tab:blooms-taxonomy}
\end{table}

\subsection{Example of Misconceptions List}
\label{app:num_misconceptions}
Following the example from the appendix \ref{app:prompt_avg_steps}, here is one of the outputs of the LLM when asked about common misconceptions for the question in the example. The prompt for the \texttt{num\_misconceptions} feature asked the LLM to generate an exhaustive, atomic list of such misconceptions, and the count was averaged over runs.

\begin{enumerate}
    \item Students may confuse 'decreasing order' with 'increasing order', leading them to reverse the correct order of the numbers.
    \item Students might incorrectly apply the property $a\sqrt{b} = \sqrt{a^2 * b}$ by forgetting to square 'a' before multiplying by 'b', calculating it as $a\sqrt{b} = \sqrt{a * b}$.
    \item Students might make arithmetic errors when squaring the number outside the square root (e.g., calculating $4^2$ as 8 instead of 16).
    \item Students might make arithmetic errors in the multiplication step within the square root (e.g., calculating $16 * 3$ as 45 instead of 48).
    \item Students may believe that the magnitude of the number outside the square root is the primary determinant of the overall value, even if the value inside the square root is smaller (e.g., incorrectly assuming that $5\sqrt{2}$ is always greater than $2\sqrt{5}$ because 5 > 2).
    \item After correctly comparing the values under the square roots, students might forget to convert them back to their original forms when selecting the final answer, leading them to choose an option with the numbers under the square roots ordered.
    \item Students might make careless copying errors when transferring the correctly ordered expressions from their intermediate steps to the final answer choice.
    \item Students may lack a strong number sense regarding the approximate values of square roots, making it difficult to estimate and compare the values without performing the conversion to a common form.
    \item Students may struggle with the concept that irrational numbers can be ordered on a number line, treating them as inherently harder to compare than integers.
    \item Students might try to apply the square root individually to each term, incorrectly stating that $a\sqrt{b} = \sqrt{a} * \sqrt{b}$.
    \item Students might not understand what a root is and how it affects the order of the numbers.
\end{enumerate}

This was performed 3 times and then averaged the number of misconceptions for each question.

\subsection{LLM Feature Extraction Prompts}
\label{app:llm_prompts}
This section details the core prompts used with Google's Gemini models (primarily \texttt{gemini-2.0-flash}, with \texttt{gemini-2.5-flash} or \texttt{gemini-2.5-pro} for more complex tasks as noted) to extract various pedagogical features. Prompts generally included instructions for chain-of-thought reasoning (within \texttt{<thinking>...</thinking>} tags, not shown in core prompts below for brevity) and a strict JSON output format. Features were typically aggregated over multiple runs (usually 1 or 3) for stability. Question-specific content like \texttt{\$\{questionTitle\}} and \texttt{\$\{optionsJson\}} were dynamically inserted.

\subsubsection{Knowledge Dimension (Feature: \texttt{Knowledge\_Dimension})}
\label{app:prompt_knowledge_dimension}
\textbf{Model Used:} \texttt{gemini-2.0-flash} (1 run)
\begin{promptbox}
Analyze the following math question. What primary type of knowledge does it assess? Choose one:

\vspace{1em}

- Factual: Recalling specific facts, definitions, formulas, or properties.

- Conceptual: Understanding relationships between concepts, interpreting information, explaining ideas, applying knowledge flexibly.

- Procedural: Executing a sequence of steps or applying a standard algorithm or method.

\vspace{1em}

Question:

\$\{questionTitle\}

\vspace{1em}

Options:

\$\{options.map(opt => `- \$\{opt\}`).join('\\n')\}

\vspace{1em}
Think step-by-step about what the student needs to know or do. Is it mainly recall, understanding connections, or following steps? Then state the single best classification.

\end{promptbox}

\subsubsection{Information Gap (Feature: \texttt{Question\_Answer\_Info\_Gap})}
\label{app:prompt_info_gap}
\textbf{Model Used:} \texttt{gemini-2.5-flash} (3 runs, mean aggregated)
\begin{promptbox}
Analyze the following math question. How much information, knowledge, or reasoning *not explicitly stated* in the question text itself is required to arrive at the correct solution? Consider implicit assumptions, required prior knowledge (beyond basic arithmetic), or necessary intermediate reasoning steps.

\vspace{1em}

Use this scale:

1 = None: The answer can be derived directly using only the information and numbers explicitly given.

2 = Low: Requires minimal prior knowledge (e.g., a common definition) or a single, obvious implicit step.

3 = Medium: Requires recall of specific formulas/theorems not given, multiple implicit reasoning steps, or interpretation of context.

4 = High: Requires significant external knowledge, complex synthesis of unstated information, or bridging substantial gaps in the problem statement.

\vspace{1em}

Question:

\$\{questionTitle\}

\vspace{1em}

Options:

\$\{options.map(opt => `- \$\{opt\}`).join('\\n')\}

\vspace{1em}

Think step-by-step about the solution process. What external knowledge (formulas, theorems, concepts) is needed? What steps rely on implicit understanding? How large is the gap between the stated information and the required solution path?
\end{promptbox}

\subsubsection{Max Expression Nesting Depth (Feature: \texttt{Max\_Expression\_Nesting\_Depth})}
\label{app:prompt_nesting_depth}
\textbf{Model Used:} \texttt{gemini-2.5-flash} (3 runs, mode aggregated)
\begin{promptbox}
Analyze the mathematical expressions in the following text. Consider nested parentheses, functions (like sqrt), fractions, exponents, etc. What is the maximum nesting depth required to parse the most complex part of any expression?

\vspace{1em}

For example:

- "2 + 3" has depth 0.

- "sqrt(5)" has depth 1.

- "3 * (4 + 5)" has depth 1.

- "sqrt(a + (b/c))" has depth 2 (division inside addition inside sqrt).

- "((a+b)\^2) / (c - d)" has depth 2.

\vspace{1em}

Text:

\$\{questionTitle\}

\vspace{1em}

Think step-by-step about the structure of each expression. Identify the most deeply nested part. Count the levels of nesting.
\end{promptbox}

\subsubsection{Most Complex Number Type (Feature: \texttt{Most\_Complex\_Number\_Type})}
\label{app:prompt_complex_number_type}
\textbf{Model Used:} \texttt{gemini-2.0-flash} (1 run)
\begin{promptbox}
Analyze the numbers involved in this math question, including those in the options.

\vspace{1em}

Consider integers, decimals, fractions, roots/irrationals (like sqrt(2) or pi), and complex numbers (like 3+2i).

\vspace{1em}

DO NOT consider abstract variables (like x) for this specific task.

\vspace{1em}

Based on the following hierarchy of *numerical* complexity (lowest to highest):

1: Integer (e.g., 5, -2, 0)

2: Decimal (e.g., 3.14, -0.5)

3: Fraction (e.g., 1/2, 3/4)

4: Root/Irrational (e.g., sqrt(3), pi)

5: Complex Number (e.g., 2i, 1+i)

\vspace{1em}

What is the HIGHEST level of numerical complexity present in the question or its options?

\vspace{1em}

Question:

\$\{questionTitle\}

\vspace{1em}

Options:

\$\{options.map(opt => `- \$\{opt\}`).join('\\n')\}

\vspace{1em}

Think step-by-step. Identify the different types of numbers present. Determine the highest level reached according to the hierarchy. Ignore variables/abstract symbols.
\end{promptbox}

\subsubsection{Has Abstract Symbols (Feature: \texttt{Has\_Abstract\_Symbols})}
\label{app:prompt_has_abstract_symbols}
\textbf{Model Used:} \texttt{gemini-2.0-flash} (1 run, result to 1/0)
\begin{promptbox}
Does the following math question or its options contain abstract symbols representing variables or unknown quantities (e.g., x, y, a, ?, width, length)? Ignore standard mathematical operators, functions, units, and concrete numbers.

\vspace{1em}

Question:

\$\{questionTitle\}

\vspace{1em}

Options:

\$\{options.map(opt => `- \$\{opt\}`).join('\\n')\}

\vspace{1em}

Think step-by-step. Look for letters or symbols used as variables.
\end{promptbox}

\subsubsection{Units Check (Feature: \texttt{Units\_Check})}
\label{app:prompt_units_check}
\textbf{Model Used:} \texttt{gemini-2.0-flash} (1 run, result to 1/0)
\begin{promptbox}
Does the following math question involve units of measurement (like meters, kg, dollars, seconds, etc.) in a way that might require attention to units or potential conversions to solve correctly?

\vspace{1em}

Question:

\$\{questionTitle\}

\vspace{1em}

Think step-by-step. Identify any units mentioned. Consider if conversions are needed.
\end{promptbox}

\subsubsection{Problem Archetype (Feature: \texttt{Problem\_Archetype})}
\label{app:prompt_problem_archetype}
\textbf{Model Used:} \texttt{gemini-2.5-pro} (1 run)
\begin{promptbox}
Classify the primary task or structure of the following math question. Choose the single best fit from this list:

\vspace{1em}

- Word Problem - Calculation

- Equation Solving

- Geometric Reasoning/Proof

- Data Interpretation

- Conceptual Definition/Understanding

- Formula Application

- Pattern Recognition

- Other

\vspace{1em}

Question:

\$\{questionTitle\}

\vspace{1em}

Think step-by-step about what the question asks the student to *do*. What is the core task?
\end{promptbox}

\subsubsection{Real-World Context Flag (Feature: \texttt{RealWorld\_Context\_Flag})}
\label{app:prompt_realworld_flag}
\textbf{Model Used:} \texttt{gemini-2.0-flash} (1 run, result to 1/0)
\begin{promptbox}
Does the context of the following math question primarily involve a real-world scenario (e.g., shopping, measuring objects, physics situations, stories about people) OR is it presented in a purely abstract mathematical way (e.g., manipulating equations, asking about number properties, abstract geometry without context)?

\vspace{1em}

Respond with either 'REALWORLD' or 'ABSTRACT'.

\vspace{1em}

Question:

\$\{questionTitle\}

\vspace{1em}

Think step-by-step about the setting and elements described in the question.
\end{promptbox}

\subsubsection{Ratio of Abstract to Concrete Symbols (Feature: \texttt{Ratio\_Abstract\_Concrete\_Symbols})}
\label{app:prompt_abstract_ratio}
\textbf{Model Used:} \texttt{gemini-2.0-flash} (1 run, calculates ratio from counts)
\begin{promptbox}
Analyze the mathematical content in the text below.

\vspace{1em}

1. Count the number of concrete numerical values (e.g., 5, 3.14, -2, 1/2).

2. Count the number of abstract symbols representing variables or unknown quantities (e.g., x, y, a, ?, width).

\vspace{1em}

Ignore standard mathematical operators (+, -, =), functions (sqrt), and units (m, kg).

\vspace{1em}

Text:

\$\{questionTitle\}

\vspace{1em}

Think step-by-step. List the concrete numbers found. List the abstract symbols found. Provide the counts.
\end{promptbox}

\subsubsection{Distractor Plausibility (Features: \texttt{LLM\_Distractor\_Plausibility\_Max}, \texttt{LLM\_Distractor\_Plausibility\_Mean})}
\label{app:prompt_distractor_plausibility}
\textbf{Model Used:} \texttt{gemini-2.5-flash} (3 runs, results aggregated for Mean and AvgMax)
\begin{promptbox}
Analyze the following math question and its options. The correct answer is '\$\{correctOption\}'.

\vspace{1em}

For each INCORRECT option provided below, rate its plausibility as a potential answer that a student might mistakenly choose.

\vspace{1em}

Use the following scale:

1 = Very Unlikely: The option is nonsensical or completely unrelated.

2 = Unlikely: The option results from a significant calculation error or conceptual misunderstanding.

3 = Possible: The option might result from a common calculation mistake or a minor misunderstanding.

4 = Plausible: The option is a common mistake or addresses a known misconception related to the question.

5 = Very Plausible: The option is highly tempting, potentially resulting from a subtle error or a very common misconception.

\vspace{1em}

Question:

\$\{questionTitle\}

\vspace{1em}

All Options:

\$\{allOptionsStr\}

\vspace{1em}

Correct Answer: \$\{correctOption\}

\vspace{1em}
Think step-by-step about why each incorrect option might be chosen or not. Consider common student errors for this type of problem. Then, provide the rating for each incorrect option.
\end{promptbox}

\subsection{Neural Network Configuration Details}
\label{app:nn_config_details}

\begin{table}[H]
    \centering
    \begin{tabular}{p{0.45\columnwidth}p{0.45\columnwidth}}
        \hline
        \textbf{Component} & \textbf{Value} \\
        \hline
        User Embedding Dimension & 8 \\
        Dense Layer Units (Input Pathways) & 32 (for numerical features), 32 (for text embeddings) \\
        Dense Layer Units (Shared) & 64, then 32 \\
        Activation Function (Hidden Layers) & ReLU \\
        L2 Regularization (All weight matrices) & 0.0005 \\
        Dropout Rate (All dropout layers) & 0.25 \\
        Optimization & Adam optimizer \\
        Learning Rate & 0.0002 (2e-4) \\
        Loss Function & Binary cross-entropy  \\
        Metrics & Accuracy, AUC \\
        Batch Size & 1024 \\
        Maximum Epochs & 60 \\
        Early Stopping Patience & 12 (monitors \texttt{val\_loss}) \\
        Restore Best Weights (Early Stopping) & True \\
        \hline
    \end{tabular}
    \caption{Neural network architecture and training hyperparameters for the best performing model.}
    \label{tab:model-details}
\end{table}

\end{document}